# New Modules for the SEDMachine to Remove Contaminations from Cosmic Rays and Non-target Light: BYECR and CONTSEP

Y.-L. Kim[1,4], M. Rigault[1], J. D. Neill[2], M. Briday[1], Y. Copin[1], J. Lezmy[1], N. Nicolas[1], R. Riddle[2], Y. Sharma[2], M. Smith[1], J. Sollerman[3], and R. Walters[2]

[1] Université de Lyon, Université Claude Bernard Lyon 1, CNRS/IN2P3, IP2I Lyon, F-69622, Villeurbanne, France; y.kim9@lancaster.ac.uk
[2] Division of Physics, Mathematics, and Astronomy, California Institute of Technology, Pasadena, CA 91125, USA
[3] The Oskar Klein Centre, Department of Astronomy, Stockholm University, AlbaNova, SE-10691 Stockholm, Sweden

*Received 2021 December 2; accepted 2022 January 28; published 2022 February 25*

## Abstract

Currently time-domain astronomy can scan the entire sky on a daily basis, discovering thousands of interesting transients every night. Classifying the ever-increasing number of new transients is one of the main challenges for the astronomical community. One solution that addresses this issue is the robotically controlled Spectral Energy Distribution Machine (SEDM) which supports the Zwicky Transient Facility (ZTF). SEDM with its pipeline PYSEDM demonstrates that real-time robotic spectroscopic classification is feasible. In an effort to improve the quality of the current SEDM data, we present here two new modules, BYECR and CONTSEP. The first removes contamination from cosmic rays, and the second removes contamination from non-target light. These new modules are part of the automated PYSEDM pipeline and fully integrated with the whole process. Employing BYECR and CONTSEP modules together automatically extracts more spectra than the current PYSEDM pipeline. Using SNID classification results, the new modules show an improvement in the classification rate and accuracy of 2.8% and 1.7%, respectively, while the strength of the cross-correlation remains the same. Improvements to the SEDM astrometry would further boost the improvement of the CONTSEP module. This kind of robotic follow-up with a fully automated pipeline has the potential to provide the spectroscopic classifications for the transients discovered by ZTF and also by the Rubin Observatory's Legacy Survey of Space and Time.

*Unified Astronomy Thesaurus concepts:* Spectroscopy (1558); Surveys (1671); Astronomy data analysis (1858); Astronomical methods (1043)

## 1. Introduction

Time-domain astronomy, the study of transients, variables, and moving objects, has reached a mature phase over the last decade. Driven by technological progress in detectors and computing power, current surveys with 1–2 m class telescopes, such as the Zwicky Transient Facility (ZTF; Bellm et al. 2019; Graham et al. 2019; Masci et al. 2019; Dekany et al. 2020), the All-Sky Automated Survey for Supernovae (Shappee et al. 2014), and the Asteroid Terrestrial Last-Alert System (Tonry et al. 2018), are able to scan the entire sky on a (near) daily basis. This leads us to observe about $O(10^3)$ interesting objects among $O(10^6)$ alerts every night (Graham et al. 2019). We expect that 10% of them are new transients that have just exploded as supernovae (SNe) or some kind of galactic or extra-galactic explosion. In the near future, the Rubin Observatory's Legacy Survey of Space and Time (LSST) with an 8 m class telescope will detect ten times more transients than current surveys (LSST Science Collaboration et al. 2009).

In that context, a classification for the ever-increasing number of new transients is one of the challenges in time domain astronomy. Accurate classification is the crucial first step in detecting new classes of astronomical objects and widening our understanding of the nature of those transients. For SN Ia cosmology, removing contamination from non-Ia types is critical to estimating unbiased cosmological parameters (e.g., Jones et al. 2017). In general, current and past surveys perform follow-up observations to obtain a spectroscopic classification using time-allocated, or non-dedicated facilities. This method of follow-up is not an efficient way to cover most of the newly discovered transients. On the other hand, a major effort is underway to develop classification methods using photometric data, which are easier to obtain (Lochner et al. 2016; Möller & de Boissière 2020, and references therein). However, photometric methods are less accurate than spectroscopic methods and not yet proven to be sufficiently reliable.

---
[4] Current address: Department of Physics, Lancaster University, Lancs LA1 4YB, UK.

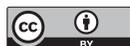 

Therefore, we need spectroscopic data acquired in an efficient way to achieve the state-of-the-art classification and to provide a training set for photometric typing.

The Spectral Energy Distribution Machine (SEDM) (Blagorodnova et al. 2018) and its pipeline PYSEDM (Rigault et al. 2019)[5] demonstrates that such spectroscopic classification is feasible in close to real-time. SEDM is a very low-resolution ($R \sim 100$) integral field unit (IFU) spectrograph, whose output is a datacube including spatially resolved spectra, with a $28 \times 28$ arcsec$^2$ field-of-view, mounted on the Palomar 60 inch telescope. The main purpose of SEDM is to classify transients discovered by ZTF. PYSEDM is a pipeline that extracts a spectrum of the SEDM targets within minutes of the exposure completing. With this pipeline, SEDM is fully automated, and hence a flux-calibrated spectrum is available in 5 minutes after the end of the exposure. The transient spectrum is classified mainly through the Supernova Identification code (SNID; Blondin & Tonry 2007) and SNIascore (Fremling et al. 2021). In this way, 15 transient spectra and their spectroscopic classifications are typically acquired every night.

As SEDM is an IFU spectrograph, it acquires simultaneously the transient and its surrounding area. This means that the SEDM datacube contains non-target contamination from the host galaxy and other sources. In addition, even though cosmic ray hits in the raw IFU CCD are rejected in the image reduction phase of the pipeline, often a cosmic ray that is coincident with a bright spaxel will pass through and impact the final spectrum. By removing the contamination from these cosmic rays and non-target light, we can obtain cleaner spectra of the SEDM targets. We expect that this target only spectrum improves the accuracy of the automated classification.

In this paper, we present two new modules that remove these types of contamination from the SEDM datacube. The first module, named BYECR, is developed to remove the cosmic ray hit contamination, while the second, named CONTSEP, removes the non-target light contamination. These two modules are presented in Section 2. Then, in Section 3 we re-extract all the SEDM data obtained from 2018 August to 2020 December and run SNID with the extracted spectra by applying the new modules. Section 4 presents the performance improvement regarding the classification rate and accuracy determined by SNID. A discussion on the next steps for the SEDM data and its classification can be found in Section 5. We summarize the paper in Section 6.

## 2. New PYSEDM Modules: BYECR and CONTSEP

Here we introduce two new modules for the PYSEDM pipeline to remove contamination from the SEDM datacube. These new modules are executed just before the "PSF spectrum extraction" step in the pipeline and are fully automated and integrated with the whole pipeline process (see Figure 1 of Rigault et al. 2019, for the flow chart of the PYSEDM pipeline).

### 2.1. BYECR: For Removing Cosmic Rays

Without correctly removing cosmic rays, spurious features will be associated with the SN leading to incorrect classifications. Cosmic rays in the SEDM CCD image are rejected by the `lacosmicx`[6] python package, after a master bias is subtracted. The `lacosmicx` is based on the L.A.COSMIC (van Dokkum 2001), which performs well on imaging or undersampled IFU data (see Husemann et al. 2012, for the detailed comparison of several cosmic ray detecting methods). As pointed out by Husemann et al. (2012), currently, cosmic-ray detection algorithms for IFU data are missing. Therefore, we implement our own algorithm for SEDM, based on a technique developed for the previous Micro-Lens Array-based IFU (e.g., the SAURON project, see Bacon et al. 2001, or the SuperNova Integral Field Instrument, see Copin 2013).

Our idea is that cosmic rays can be removed by utilizing spatial information in the SEDM datacube. Since we require spatial information, our method will be applied at the datacube level, where the spatial transformation is defined, rather than at the raw CCD image. As illustrated in the left panel of Figure 1, we first perform a normalization of each spectrum along the spectral axis as follows. For each voxel (volume pixel which refers to a volume element of a datacube), we divide its flux by the average flux in the $-10$ to $+10$ spectral bin window of the voxel spectrum. Then, the BYECR module compares the flux of a test voxel (black) with the mean flux estimated from its 6 spatial neighbors (6 different colors). When the test flux shows a $5\sigma$ excess in comparison to the mean flux, the BYECR module identifies this test voxel as contaminated by a cosmic ray. The cosmic ray contaminated voxel is removed (the flux value is converted to "*NaN*") and ignored when the 2D PSF modeling for the SEDM datacube is performed via PYSEDM. From this, we can obtain a more robust PSF model and therefore a cleaner spectrum, as shown in the right panel of Figure 1.

### 2.2. CONTSEP: For Removing Non-target Contamination

The current PYSEDM, documented in Rigault et al. (2019), fits a 3D PSF that has a fixed angular elliptical size. Hence, the extracted target spectrum contains light from neighboring sources. For an SN, this is mostly from its host galaxy. Therefore, some extra pre-processing is required to ensure that the transient spectrum is as clean as possible for more accurate classification (see Blondin & Tonry 2007, for more discussions about this). To mitigate this issue, we have developed the CONTSEP module that removes spaxels dominated by non-target sources.

---

[5] https://github.com/MickaelRigault/pysedm

[6] https://github.com/cmccully/lacosmicx

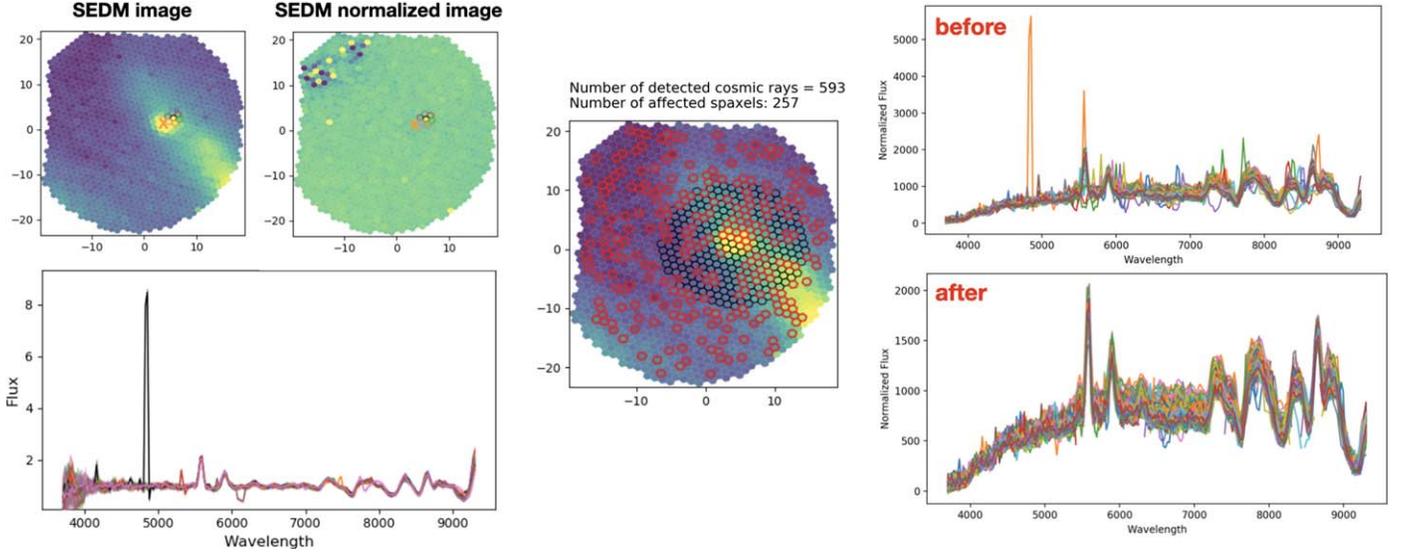

**Figure 1.** Example of how the BYECR module finds cosmic rays in the SEDM datacube of SN ZTF19acdabep. Top left: Illustration of the SEDM image and its normalized image obtained by the BYECR module. The orange cross indicates the center of the target. The black spaxel in the center of color spaxels is the cosmic ray test spaxel and the six different color spaxels are its neighbors. The BYECR module calculates the mean flux of the six neighbors in each of the 220 wavelength bins (i.e., voxel). Bottom left: Spectra of the cosmic ray test spaxel (black) and its six neighbors selected above. There is a strong cosmic ray around 4900 Å. Middle: Detected cosmic rays by the BYECR module (red spaxels). For this sample, there are 593 cosmic rays in 257 spaxels. Black spaxels show the default aperture size of PYSEDM, from which the target spectra will be extracted. Right: Extracted spectra before (upper) and after removing cosmic rays (lower).

The basic idea is to find the faintest contour that separates the target and other sources. As illustrated in the left panel of Figure 2, first we take an *r*-band cut-off image from the Pan-STARRS survey (hereafter PS; Chambers et al. 2016) centered at the target coordinates. We then insert a pseudo-target which has 16.0 mag on the PS image based on the SEDM astrometry. The value of 16.0 mag was selected after several tests during the development phase and produces the best results. It corresponds to the best balance between bright enough to be clearly visible within any host galaxy, and faint enough not to outshine other sources. Future development could test variable magnitudes depending on the actual target magnitude when triggered by SEDM.

Because the PS data has magnitude information, the CONTSEP module can draw isomagnitude contours on the PS image and project them onto the SEDM image. From the isomagnitude contours, the module finds the faintest contour which separates the target and other sources. Then, the CONTSEP module selects spaxels inside the target contour for the target (red in Figure 2), and the other sources' contour for non-target sources (black). The selected non-target sources' spaxels are removed from the default PYSEDM aperture when we extract the target spectrum using the 3D PSF procedure, described in Rigault et al. (2019). As a result, the final spectrum processed by CONTSEP is expected to be cleaner than the default PYSEDM spectrum as shown in the right bottom right panel of Figure 2.

### 3. Re-extraction and Run SNID

In order to test our new modules, we have re-extracted all the SEDM spectra obtained from 2018 August to 2020 December[7] with different module combinations:

*pysedm*: a default PYSEDM,
*byecr*: PYSEDM + BYECR,
*contsep*: PYSEDM + CONTSEP,
*byecont*: PYSEDM + BYECR + CONTSEP

In total, 6784 spectra for 4630 individual transients were re-extracted (Table 1). When the CONTSEP module is applied, the number of extracted spectra is 128 (∼2%) fewer than when not applied. Among those not-extracted spectra, 65 are due to an SEDM pointing failure, 43 are a PS image issue (e.g., saturation), 15 are aligned with the host core (a strong host or active galactic nucleus (AGN) case), and 5 are CONTSEP failures.

Next, we apply a cut based on the signal-to-noise ratio (S/N) of the extracted SEDM spectra. We calculate S/N (per ∼25 Å bin) in the wavelength range between 4000 and 8000 Å that will be adopted for SNID, and take the median. After examining the spectra, we select S/N > 3 as the cut criterion where

---
[7] We note that spectra from 2019 May 14 are removed due to a technical issue with the instrument.

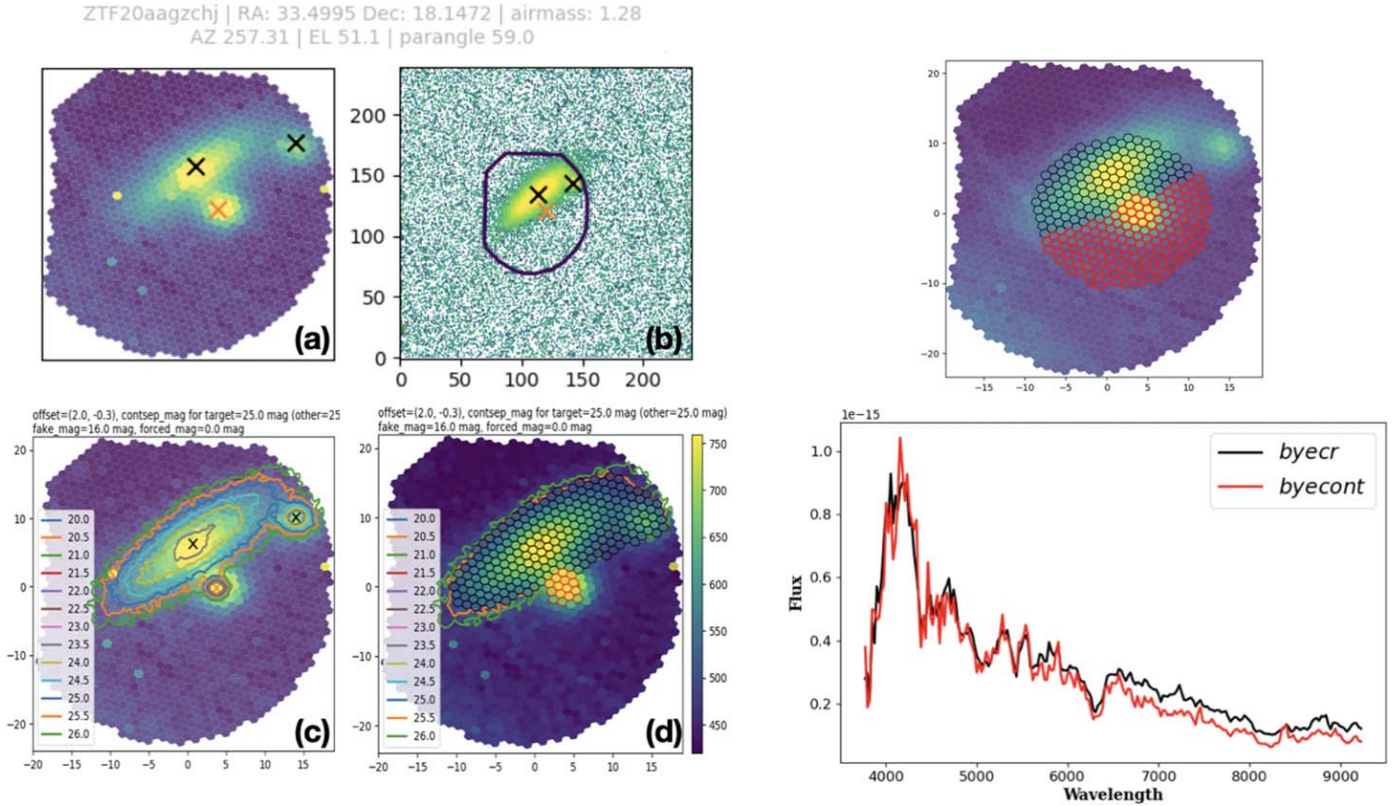

**Figure 2.** Example of how the CONTSEP module removes non-target contaminations in the SEDM data of SN Ia ZTF20aagzchj. Left: Panel (a) shows the SEDM image. The orange cross indicates the center of the target, while two black crosses are the non-target sources: the left one is a host galaxy and the right one is an infrared source according to the NASA Extragalactic Database. Panel (b) shows the $r$-band PS cut-off image, centered at the target coordinate. The figure in the middle represents the SEDM footprint for this target. In Panel (c), the CONTSEP module draws isomagnitude contours based on the PS optical image, and finds that 25.0 mag (cyan line; CONTSEP mag) is the faintest contour to separate the target from other sources. The CONTSEP module selects spaxels for the target (red) and other sources (black) based on the CONTSEP mag (panel (d)). We note that the offset = (2.0, −0.3) in the unit of spaxel has been applied manually for this target (see Section 5.1). Right: Selected other spaxels are removed from the default PYSEDM aperture (black + red) when we extract the target spectrum. Therefore, the spectra only in red spaxels are extracted. In the bottom panel, extracted flux-calibrated spectra with (black; *byecr*) and without (red: *byecont*) contamination light are presented.

**Table 1**
The Sample Size of Each Extraction Method Obtained from 2018 August to 2020 December

|         | All Spectra | Extracted |         | S/N > 3 |         |                   |
|---------|-------------|-----------|---------|---------|---------|-------------------|
|         |             | Spectra   | % diff. | Spectra | % diff. | Mean (median) S/N |
| *pysedm*  | 6784        | 6784      | ⋯       | 5800    | ⋯       | 13.39 (9.67)      |
| *byecr*   |             | 6784      | 0%      | 5823    | +0.4%   | 13.10 (9.42)      |
| *contsep* |             | 6656      | −1.9%   | 5835    | +0.6%   | 12.54 (9.09)      |
| *byecont* |             | 6656      | −1.9%   | 5852    | +0.9%   | 12.43 (8.98)      |

**Note.** *% diff.* is the relative fractional percent difference compared to the default *pysedm* method.

spectral features are shown. The number of final spectra with the mean and median S/N for each extracted method is presented in the last column of Table 1. Because *pysedm* includes all the light from contamination, it has the highest value of S/N, while since *byecont* removes all the contamination, it has the lowest. Interestingly, the final number of spectra

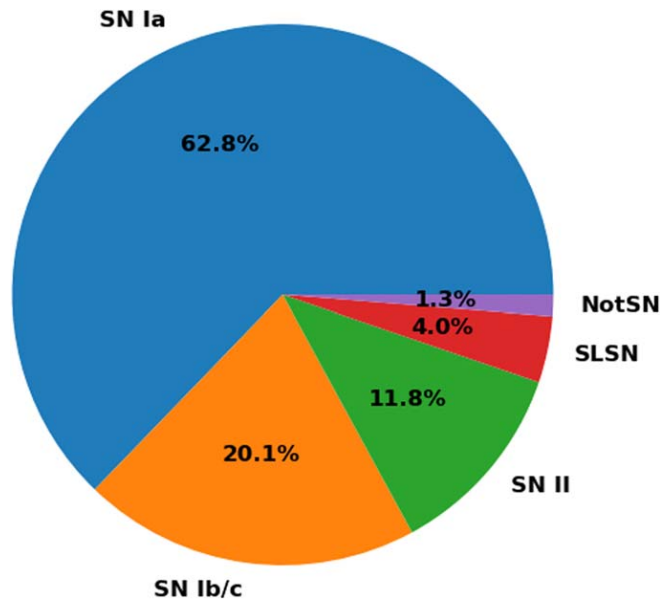

**Figure 3.** The fraction of SNID template spectra for each type used.

## 4. SNID Results

We examine the SNID results in various aspects to test the performance of the new modules. Table 2 summarizes the results from each extraction method.

### 4.1. Classification Rate

First, we calculate how many SEDM spectra are classified by SNID. We calculate the rate based on all the 6784 SEDM spectra. As shown in the columns between second and fourth of the table, each extraction method has a different classification rate. Removing cosmic ray hits (*byecr* method) improves the rate by 1.6% (165 more spectra) compared to the default *pysedm* method. Removing non-target contamination shows a 0.5% (33 more spectra) improvement. Combining the two new modules improves the result with 2.8% (188 more spectra) in total compared to the default *pysedm* method.

### 4.2. rlap

The SNID *rlap* value is a measurement of the strength of the correlation between a template spectrum and the input spectrum (Blondin & Tonry 2007). The distribution and the mean (median) of *rlap* of the best-matching template are presented in Figure 4 and the fifth column of Table 2, respectively. Their distributions and values are similar, however we observe a slight improvement when applying the new modules. Removing cosmic ray hits improves the mean *rlap* by 0.29 (∼3%). However, removing non-target contamination produces an *rlap* value similar to that of *pysedm*. This is because the CONTSEP module requires an accurate astrometric solution, which sometimes fails in the automated pipeline (see Section 5.1 for more discussion about this issue). In addition, the CONTSEP module does not work well when the target is aligned with the core of the host galaxy.

### 4.3. Accuracy

In order to test the classification accuracy determined by SEDM with SNID, we compare our final classifications to those made on the ZTF Bright Transient Survey (BTS) sample (Fremling et al. 2020). The BTS sample was selected because SEDM was employed as the primary classification instrument and their classification of the transients have been carefully vetted by humans and supplemented by spectra from larger telescopes (Fremling et al. 2020).

For calculating the accuracy, we first match our sample to the BTS sample. Then we construct a confusion matrix for each extraction method, considering the BTS classification as the *True label* and the SEDM with SNID classification as the *predicted label* (Figure 5). Based on this confusion matrix, we is higher when the CONTSEP module is applied (35 and 29 more for *contsep* and *byecont* methods, respectively), even though it extracted 128 (∼2%) fewer spectra at first. Comparing to the default *pysedm* method, *byecont*, which includes both new modules, provides 52 (∼1%) more spectra. We next ran SNID on this final sample of spectra to compare classification results.

When running SNID, we used SEDM spectra in the wavelength range between 4000 and 8000 Å, where the quantum efficiency of the SEDM CCD is over 60% and includes most of the important spectral features of the transients. We applied $rlap_{min} \geqslant 5$ for a classification criteria, as suggested by Blondin & Tonry (2007). The SNID templates used in the paper include a training set for DASH (Deep Automated Supernova and Host Classifier; Muthukrishna et al. 2019), as well as SN IIP templates from Gutiérrez et al. (2017), SLSN-Ic from Liu et al. (2017), and several SLSN-I, SLSN-IIn and TDE added by J.D.Neill. The DASH training set combines the spectra from the SNID Templates 2.0, SN Ib/c from Liu & Modjaz (2014), Liu et al. (2016), and Modjaz et al. (2016), and the Berkeley SN Ia program v7.0 (Silverman et al. 2012). It also removed spectra where the date of maximum was unknown. In total, for SN templates, 5170 unique spectra from 460 individual SNe, and for non-SN templates, 67 spectra from 23 stellar objects are used for our analysis. The resulting SNID template type distribution consisted of 3288 spectra from 312 SNe Ia, 1055 spectra from 80 SNe Ib/c, 620 spectra from 33 SNe II, 207 spectra from 35 SLSNe, 29 spectra from 7 TDE, 15 spectra from 3 LBVs, 11 spectra from 11 galaxies, 11 spectra of M-stars, and 1 spectrum from 1 AGN. The fraction of spectra for each type is illustrated in Figure 3.

Table 2
SNID ($rlap \geqslant 5$) Results of Each Extraction Method

|  | Classified | | | rlap | | Accuracy | | | |
| --- | --- | --- | --- | --- | --- | --- | --- | --- | --- |
|  | Spectra | Rate | % diff. | Mean (median) | % diff. | all Types | % diff. | only Ia | % diff. |
| *pysedm* | 4986 | 73.5% | ⋯ | 8.92 (7.70) | ⋯ (⋯) | 58.3% | ⋯ | 75.9% | ⋯ |
| *byecr* | 5151 | 75.9% | +1.6% | 9.21 (7.97) | +3.3% (+3.5%) | 58.9% | +0.6% | 75.8% | −0.1% |
| *contsep* | 5019 | 74.0% | +0.5% | 8.91 (7.72) | −0.1% (+0.3%) | 59.8% | +1.5% | 77.0% | +1.1% |
| *byecont* | 5174 | 76.3% | +2.8% | 9.23 (8.02) | +3.5% (+4.2%) | 60.0% | +1.7% | 76.4% | +0.5% |

**Note.** *Classified rate* is the fraction of SNID classified spectra over all SEDM 6784 spectra. *% diff.* is the relative fractional percent difference compared to the default *pysedm* method.

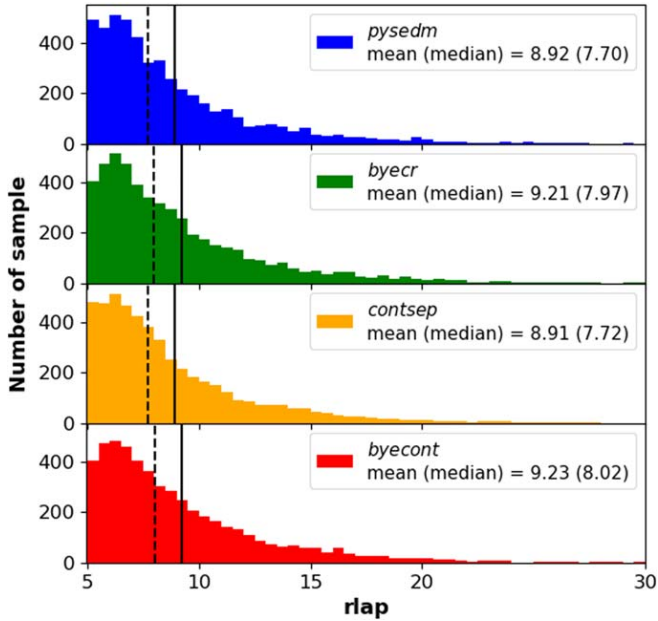

**Figure 4.** SNID *rlap* distribution for each extraction method. The solid lines indicate the mean and the dotted lines show the median of *rlap*.

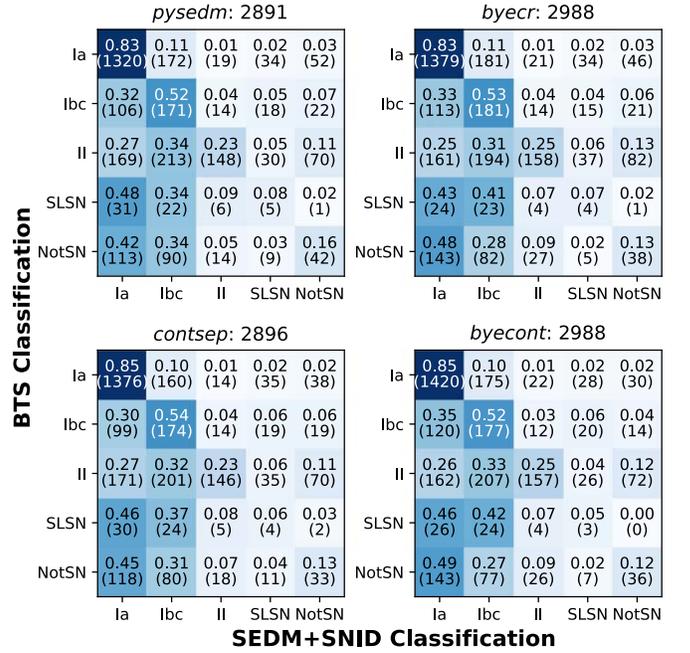

**Figure 5.** Normalized confusion matrix for SNID ($rlap \geqslant 5$) classifications with SEDM spectra extracted by each extraction method. The matrix shows the fraction (absolute number) of each BTS classified type (the True label) that was classified from SEDM data with SNID (the Predicted label). The diagonal elements indicate the data correctly classified, while off-diagonal elements are the data that are misclassified by SEDM with SNID. The title of each subplot shows the extraction method along with the number of matched spectra with the BTS sample.

calculate the overall accuracy as

$$\text{Overall Accuracy} = \frac{\text{Total number of true positive}}{\text{Total number of spectra}}. \quad (1)$$

The total number of the matched samples for each extraction method is indicated in the title of each subplot in Figure 5. Our classification accuracy for all types and only for SNe Ia is presented between the seventh and the last columns of Table 2. Removing contamination from cosmic rays and non-target sources improves the accuracy, up to ∼2%.

When looking at the confusion matrix in Figure 5, SEDM with SNID is able to classify ∼85% of the SNe Ia but with the false positive rate of ∼33%. We consider that this is because most of the samples (∼61%) across all types are classified as SNe Ia. For other types, determining the type is more uncertain.

Interestingly, about half of BTS SLSN and NotSN types are classified as SNe Ia by SEDM with SNID. We would say that it is due to the SNID template: SLSNe and peculiar transients are not really represented, while SNe Ia are over-represented. Therefore, when SNID is faced with a random spectrum, especially if it is low S/N, it will often match to an SN Ia, because there are so many Ia's that sometimes shows randomly a good match.

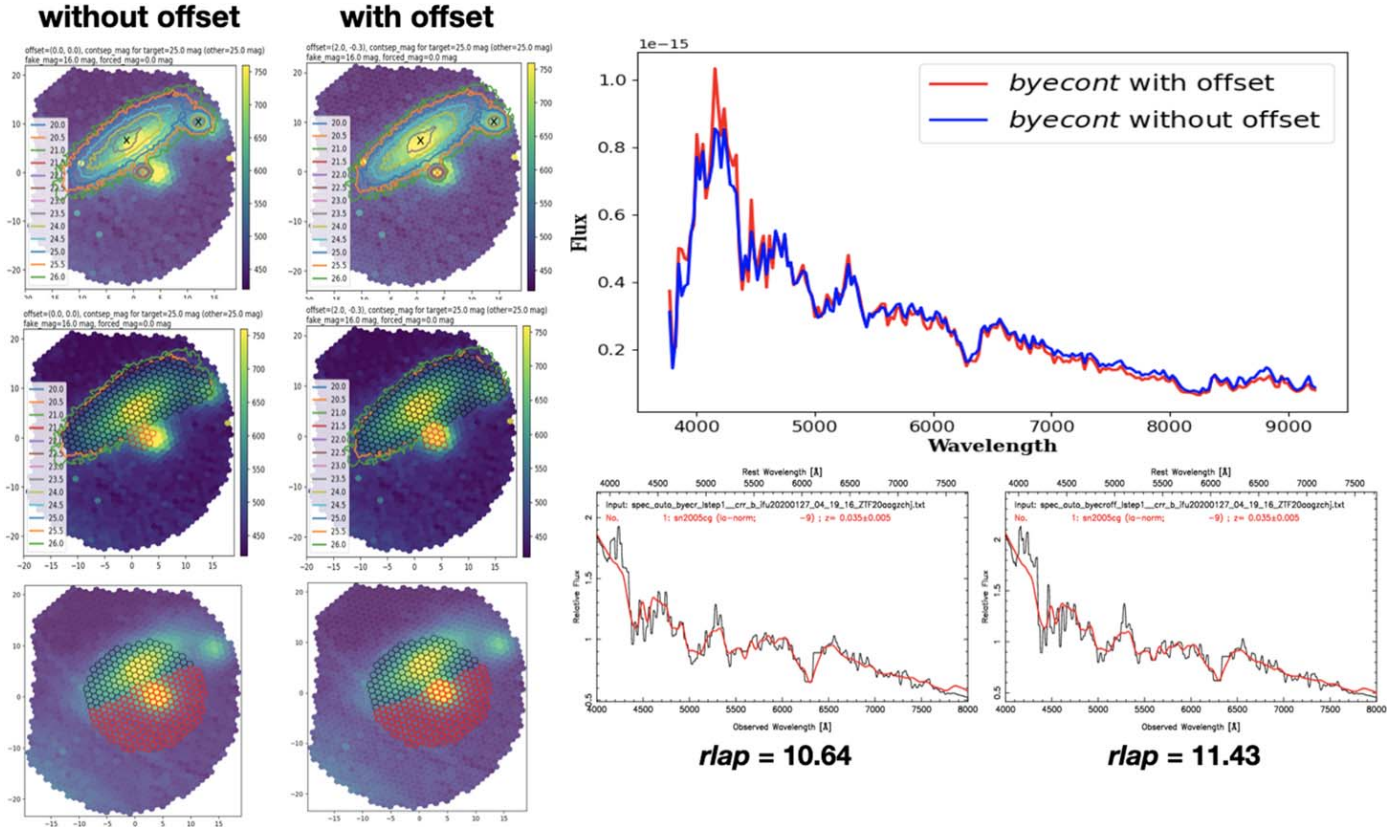

**Figure 6.** Impact of the SEDM astrometry on the CONTSEP module showing the same target as in Figure 2. Left: The location of the target is marked with the orange cross surrounded by circular isocontours, and black crosses indicate non-target sources. The isomagnitude contours are drawn based on the PS image with the astrometry information from SEDM. As shown in the left column, which shows the default automatic PYSEDM pipeline (without offset), the locations of sources are mis-aligned between the PS and SEDM images. When we set the offset = (2.0, −0.3) in the unit of spaxels, the source locations are well aligned (right column). The following figures show the selected non-target spaxels by CONTSEP module (black) and the aperture of PYSEDM (red). When there is no offset, the PYSEDM aperture has more non-target spaxels. This difference leads to different shapes of spectra and a different SNID result (right panel).

## 5. Next Steps for SEDM and its Classifications

In this section, we discuss several possibilities to further improve SEDM data and its classification accuracy.

### 5.1. Accurate Astrometry for the CONTSEP Module

The CONTSEP module works with the PS image based on the target coordinate from SEDM (see Figure 2). This means that the astrometry solution from SEDM plays an important role when we utilize the CONTSEP module. For example, for the target in Figure 6, we manually applied the offset of $\Delta X = +2.0$, $\Delta Y = −0.3$ spaxels (corresponding to $+1.5''$, $−0.4''$). This offset aligns the target and the other sources in the SEDM image with those in the reference PS image. However, there is currently no way to apply this offset automatically, even though it critically improves the results from the CONTSEP module.

As shown in the left column of Figure 6, when the SEDM astrometry is not accurate and the offset is not provided, the positions of sources in the SEDM image are not well-aligned with those in the PS image. Hence, the isomagnitude contours derived from the PS image cannot be placed accurately on the SEDM image. With poorly aligned contours, the CONTSEP module automatically selects contaminated spaxels. Consequently, this affects the shape of the target spectra and thus the classification determined by SNID (right column of Figure 6). SNID returns an *rlap* value of 10.64 without the offset and 11.43 with the offset. From this example, we can (partially) explain the relatively low value of *rlap* when we use the *contsep* method without the offset presented in Table 2.

This astrometry offset is most likely due to bad weather that affects the derivation of an accurate astrometric solution from "the guider-astrometry position" method currently used in SEDM, as detailed in Rigault et al. (2019). Briefly, SEDM uses the "Rainbow Camera" for guiding during IFU observations (see Blagorodnova et al. 2018, for the Rainbow Camera in detail). The camera has a 13 × 13 arcmin$^2$ field-of-view split into four filter quadrants. Each quadrant has one of the four filters ($u$, $g$, $r$, $i$) and also has the IFU pick-off prism mounted in the center. The position of the IFU pick-off prism is fixed, so that one can project a WCS solution estimated from the guider image onto the IFU. For this, the PYSEDM pipeline creates a median stack of all the guider images. We suspect that this median stacking from four different filters can cause inaccurate astrometry. Work on improvements to the SEDM astrometry with the automatic offset estimator is ongoing.

### 5.2. Removing Strong Host Contamination

The CONTSEP module cannot remove host galaxy contamination when a target is located at/near the center of the galaxy (about 2 spaxels or ~1.5″ away). Based on Fremling et al. (2020, their Figure 6), which presents the cumulative distribution of the angular offsets in arcsec between SNe and their hosts, we expect that ~20% of all the objects would fall within this radius from the nucleus of its host. In this case, the host contamination is stronger than when the target is at a larger distance. To remove this strong host contamination, work on modeling host spectra and the projection onto the SEDM datacube is ongoing (J. Lezmy et al. 2022, in preparation). When this work is applied to the SEDM data, we can separately obtain SN and host spectra to improve the SN classification and also to measure the host galaxy properties.

### 5.3. More SNID Templates, Especially for other SN Types

SEDM with SNID finds the majority of targets to be SNe Ia (see Figure 5). Classifying other SN types is more uncertain. Even though we supplemented the templates with other non-SNIa types, still 62.8% of all templates are SNe Ia templates (Figure 3). This imbalance of templates biases the classifications toward SNe Ia, regardless of their true type, when the spectra have a low S/N value. This issue, called a "type attractor", is also discussed by previous studies (e.g., Blondin & Tonry 2007; Silverman et al. 2012; Muthukrishna et al. 2019). In order to improve the utility of the template-based classification methods, such as SNID and Superfit (Howell et al. 2005), we require more templates particularly for other types of transients. We expect that SEDM spectra, which observes ~15 transients per night, can contribute templates to help address this issue.

### 5.4. Deep Learning Classifier

There are advances in the use of deep learning techniques to determine the type of newly discovered transients (Charnock & Moss 2017; Moss 2018). Among them, few attempts use spectroscopic classification (e.g., Muthukrishna et al. 2019;

Fremling et al. 2021). However, SNIascore (Fremling et al. 2021) was designed specifically for spectroscopic classification focusing on SNe Ia with a training set using SEDM data. It reaches the classification accuracy up to 90% while maintaining a very low false positive rate (<0.6%). This high performance is achieved even though SNIascore was trained on SEDM data extracted with the *pysedm* method. Hence, we expect that when it will be trained with cleaner data extracted with *byecont*, SNIascore will show even better performance. Work on this and a more sophisticated deep learning method to classify also other types than SNe Ia is ongoing (Y. Sharma et al. 2022, in preparation).

## 6. Summary and Plan

In this paper, we have introduced two new modules for the PYSEDM pipeline: the BYECR and CONTSEP modules. The purpose of the new modules is to remove cosmic rays and non-target contamination, respectively, for obtaining cleaner SEDM data. In order to test the performance improvement, we re-extracted all the SEDM data obtained from 2018 August to 2020 December, which consists of 6784 spectra for 4630 individual transients. Employing the new modules allows the extraction of up to 52 (~1%) more spectra than the current PYSEDM pipeline. Then, we ran SNID on the extracted spectra from different extraction methods as a comparison. When we use both modules (*byecont* extraction method), the classification rate is improved by 2.8% (188 spectra) and for the classification accuracy a 1.7% improvement is observed, compared to the current PYSEDM pipeline. However, the average (median) *rlap* value remains the same, showing a marginal improvement by 0.31 (0.32). Work on improving SEDM data and thus the spectroscopic classification is ongoing by improving the astrometric accuracy of SEDM and by removing strong host contamination in SEDM data. Along with this, more robust spectroscopic classification methods will be required through adding more templates for SNID and applying advanced deep leaning techniques.

The classification of SEDM data using SNID is currently annotated and reported to registered clients via the GROWTH Marshal (Kasliwal et al. 2019) and Fritz (Duev et al. 2019; van der Walt et al. 2019).[8] Only SNe Ia classified by SNIascore (threshold of 0.9) are reported to the Transient Name Server (TNS)[9] (Fremling et al. 2021). From the above efforts to improve the SEDM classification, we are looking forward to reporting all types of transients classified from SEDM data to TNS. Furthermore, we plan to launch SEDMv2 mounted on the Kitt Peak National Observatory 2.1 m telescope. We expect that SEDMv2 will observe a similar number (15 per night) of but fainter targets (down to ~19.0 mag) as SEDM. Robotic follow-up with SEDM and SEDMv2 will provide spectroscopic data for transients brighter than 19.0 mag, discovered by ZTF and also by the Rubin Observatory's LSST.

---

[8] https://github.com/fritz-marshal/fritz
[9] https://wis-tns.org

We thank the referee for his/her careful reading of the manuscript and many helpful comments. This project has received funding from the European Research Council (ERC) under the European Union's Horizon 2020 research and innovation program (grant agreement No. 759194—USNAC).

The SED Machine is based upon work supported by the National Science Foundation under grant No. 1106171.

Based on observations obtained with the Samuel Oschin Telescope 48 inch and the 60 inch Telescope at the Palomar Observatory as part of the Zwicky Transient Facility project. ZTF is supported by the National Science Foundation under grant No. AST-1440341 and a collaboration including Caltech, IPAC, the Weizmann Institute for Science, the Oskar Klein Center at Stockholm University, the University of Maryland, the University of Washington, Deutsches Elektronen-Synchrotron and Humboldt University, Los Alamos National Laboratories, the TANGO Consortium of Taiwan, the University of Wisconsin at Milwaukee, and Lawrence Berkeley National Laboratories. Operations are conducted by COO, IPAC, and UW.

This work made use of the data products generated by the NYU SN group, and released under DOI:10.5281/zenodo.58766, available at https://github.com/nyusngroup/SESNtemple/. The ztfquery (Rigault 2018) code is used to access SED Machine data.

### ORCID iDs

Y.-L. Kim https://orcid.org/0000-0002-1031-0796
M. Rigault https://orcid.org/0000-0002-8121-2560
J. D. Neill https://orcid.org/0000-0002-0466-1119
Y. Copin https://orcid.org/0000-0002-5317-7518
N. Nicolas https://orcid.org/0000-0002-2681-6580
R. Riddle https://orcid.org/0000-0002-0387-370X
M. Smith https://orcid.org/0000-0002-3321-1432
J. Sollerman https://orcid.org/0000-0003-1546-6615